\documentclass[fleqn,twoside]{article}
\usepackage{gc}
\usepackage{graphicx}

\heads{V.I. Dokuchaev, Yu.N. Eroshenko and S.G. Rubin}
      {Quasars formation around clusters of primordial black holes}

\begin{document}
\twocolumn[
\Arthead{0}{2004}{0 (00)}{0}{00}

\Title{Quasars formation around clusters of primordial black holes}

   \Authors{V.I. Dokuchaev\foom 1, Yu.N. Eroshenko\foom 2 }           
          {and S.G. Rubin\foom 3 }   
           {Institute for Nuclear Research of the Russian Academy of Sciences,            
            Moscow, Russia}      
           {Moscow State Engineering Physics Institute, Moscow,
 Russia; \\ Center for Cosmoparticle Physics "Cosmion", Moscow, Russia}      

\Abstract{We propose the model of first quasars formation around the cluster
of primordial black holes (PBHs). It is supposed, that mass fraction of the
universe $\sim10^{-3}$ is composed of the compact clusters of PBHs, produced
during the phase transitions in the early universe. The clusters of PBHs
become the centers of dark matter condensation. As a result, the galaxies
with massive central black holes are formed. In the process of galaxies
formation the central black holes are growing due to accretion. This
accretion is acompaned by the early quasar activity.}


]  
\email 1 {dokuchaev@inr.npd.ac.ru}
\email 2 {erosh@inr.npd.ac.ru}
\email 3 {sergeirubin@list.ru}


\section{Introduction}

The discovery of distant quasars with the redshifts $z>6$ in the Sloan Digital
Sky Survey \cite{z6} provides new questions in the problem of galaxy
formation. The maximum observed quasar redshift $z=6.41$ corresponds to
corresponds to accretion on the black hole (BH) with the mass
$3\cdot10^9M_{\odot}$ \cite{will03}. The early formation of BHs with masses
$\sim10^9M_{\odot}$ puts serious difficulties on the BH formation scenario
due to the dissipative evolution of star clusters \cite{dokrev}, supermassive
or gaseous disks \cite{el2}.  In addition it is dfficult to reconcile the
shape of the quasar redshift distribution with the hypotheses of slow
continuous BH growth in the processes of their merging or accretion
\cite{vester02, diet02}.  For these reasons the scenario of PBHs origin
\cite{zeld67, carr75} becomes more attractive.  These PBHs can be the centers
of baryon \cite{Ryan} and dark matter (DM) \cite{DokEroPAZH} condensations
into the growing galaxies.  

The new effective mechanism of PBHs formation was developed in 
\cite{Ru1,Ru2,KR04}.  This model predicts the specific 
cluster structure and properties of the forming PBHs \cite{KR04}, \cite{Ru2a}. 
In the current paper we will use this model for describing the formation of
galaxies around the clusters of PBHs.  It is the clusters of PBH who could
play the role of the initial density perturbations. As the basic example, a
scalar field with the tilted Mexican hat potential was accepted.  Properties
of PBHs clusters appear to be strongly dependent on the initial phase of the
scalar field.  In addition they strongly depend on the tilt of the potential
and the scale of symmetry breaking $f$ at the beginning of the inflation. 
Here we elaborate the dynamics of DM coupled with a PBH cluster by gravity. 
It is shown that the galaxies could be formed in this case even in the
absence of the primordial fluctuations in the DM.  We will use here the same
values of parameters as in the \cite{Ru1,Ru2}, which are in the agreement
with observations.

\begin{figure}[t] 
\includegraphics[angle=0,width=0.45\textwidth]{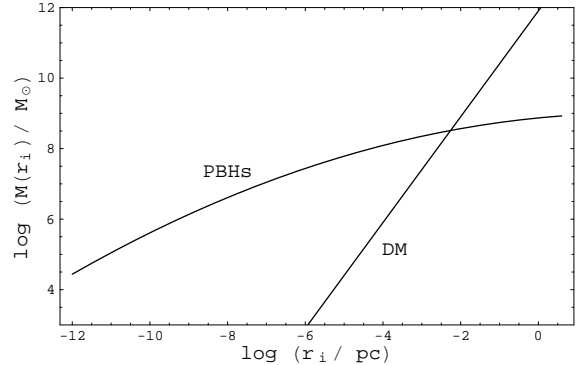}
\caption{\label{massprof1} The initial mass profile $M_h(r_i)$ of PBHs cluster
and the DM mass profile $M_{\rm DM}(r_i)$.  } 
\end{figure}

The initial mass profile $M_h(r_i)$ of PBHs cluster is calculated here
according to \cite{Ru2a} and has the specific form presented in the
Fig.~\ref{massprof1}.  For comparison the mass of DM $M_{\rm DM}(r_i)$ inside
the same spherical shells are shown.  Radius $r_i$ denotes the physical size
at the moment $t_i$ and the temperature $T_i$, when the corresponding sphere
are crossing the cosmological horizon.  Any physical size at the
temperature $T_i$ is smaller than that in the recent epoch at $T_0/T_i$ times,
where $T_0=2.7$~K.  Note that shells in the Fig.~\ref{massprof1} are taken at
different moments $t_i$. Respectively the shown mass of uniformly distributed
DM does not follow to the law $M_{\rm DM}\propto r^3$ as it must be for fixed
time case.

The mass distribution of PBHs is strongly distorted after their formation due
to the fractal structure of the cluster. Indeed, an each PBH in the cluster
is accompanied by the subcluster of smaller mass PBHs. The much more massive
PBH is formed after the corresponding gravitational radius radius crosses the
horizon. The density in the local center is so high that a lot of PBHs appear
to be inside their common gravitational radius $r_g=2GM/c^2$. The total mass
appears inside the horizon is $4.3\cdot10^7M_{\odot}$. So the whole range of
masses and radii shown in the Fig.~\ref{massprof1} is not realized. As a
result we obtain the initial mass distribution of PBHs with the much more
massive PBHs than it could be expected. In fact, the satellite PBHs of
smaller masses increase this value in several orders of magnitude. The next
Sections are devoted to the investigation of mutual evolution of DM
(uniformly distributed from the beginning) and the cluster of PBHs with distribution
according to the Fig.~\ref{massprof1}. Note that the PBHs themselves
contribute to the DM, but we don't consider them as the main DM part.

There are several stages of PBHs and galaxies formation:
(i) The formation of the closed walls of scalar field just after
inflation and their collapse into the cluster of PBHs according to
\cite{Ru2a}. Formation of the most massive BH in the center of the
cluster after the horizon crossing. (ii) The detaching of the central
dense region of the PBHs cluster from the cosmological expansion and its
virialization. Many of surrounding small BHs merge with a central most massive
BH and increase its mass. (iii). The quasars ignition due to accretion onto
the central BH. (iv) Detaching of the outer cluster region where DM dominates
from the cosmological expansion and further galaxy growth. Stop of the galaxy
growth due to interaction with the surrounding DM
fluctuations originated from the inflation. (v) The gas cooling and star
formation.

There are definite astrophysical limitations on the number and mass of PBHs:
The universe age limit gives for PBH cosmological density parameter
$\Omega_h\le1$. The possibility of tidal destruction of globular clusters by
the PBHs gives the PBH mass limit $M_h\le10^4M_{\odot}$ if these PBHs provide
the major contribution to the DM \cite{Moore93}. The contribution of accreted
PBHs into the background radiations gives approximately $\Omega_h\le10^{-2}$
for $M_h\sim10^5M_ {\odot}$ \cite{Carr79}. The limit $\Omega_h<0.01$ for
$10^6M_ {\odot}<M_h<10^8M_{\odot}$ is obtained from the VLBI observations of
the lensing of compact radio--sources \cite{Wilk01}. In this paper we
consider the case $\Omega_h\sim10^{-3}$ which is in accordance with all
aforementioned limits.

In the following the subscript '0' marks the values of a current time $t_0$,
'eq' corresponds to the time $t_{\rm eq}$ of matter--radiation equality, and
'i' to the time of horizon crossing respectively. We consider the flat
cosmological model with the density parameters $\Omega_{m,0}=0.3$,
$\Omega_{\Lambda,0}=0.66$, $\Omega_{b,0}h^2=0.02$ and $h=0.7$.

\section{Gravitational dynamics of cluster}

Let us consider the gravitational dynamics of  PBHs cluster and the DM. The
results of this section are applicable as for the inner part of the cluster,
composed mainly of PBHs and collapsing at radiation dominated stage, and for
outer regions of the cluster, where the DM is the main dynamical component.
The later regions are detached from the cosmological expansion at the matter
dominated epoch. The spherical symmetry is supposed. This approximation is
rather good for the inner regions because the considered density fluctuations
have a large amplitude in comparison with the standard inflationary generated
fluctuations.

Consider a spherically symmetric system with the radius $r<ct$, consisting of
PBHs with the total mass $M_h$ inside the radius $r$, the radiation density
$\rho_r$, the DM density $\rho_{\rm DM}$ and the cosmological  
$\Lambda$--term  density $\rho_{\Lambda}$. The radiation density (and
obviously the density of $\Lambda$-term) is homogenous. Therefore, the
fluctuations induced by the PBHs are of the type of entropy fluctuations.
Because the scale under consideration is less then the cosmological horizon
scale we use the Newtonian gravity but take into account the prescription of
\cite{TolMc30} to treat the gravitation of homogenous relativistic components
$\rho\to\rho+3p/c^2$. the evolution of spherical shells obeys the following
equation
\begin{equation}
\frac{d^2r}{dt^2}=-\frac{G(M_h+M_{\rm DM})}{r^2}-\frac{8\pi G\rho_r
r}{3}+\frac{8\pi G\rho_{\Lambda} r}{3} \label{d2rdt1}
\end{equation}
with the approximate initial conditions at the moment $t_i$: $\dot
r=-Hr$, $r(t_i)=r_i$. During the derivation of Eq.~(\ref{d2rdt1}) it was taken
into account, that $\varepsilon_r+3p_r=2\varepsilon_r$,
$\varepsilon_{\Lambda}+3p_{\Lambda}=-2\varepsilon_{\Lambda}$. In
the parametrization $r=\xi a(t)b(t)$, the $\xi$ is the comoving length, $a(t)$
is the scale factor of the universe normalized to the present moment $t_0$ as
$a(t_0)=1$ and the function $b(t)$ characterizes the deflection of the
cosmological expansion from Hubble law. The $\xi$ connected to the mass of DM
inside considered spherical volume (excluding BHs mass) by the relation
$M_{\rm DM}=(4\pi/3)\rho_{\rm DM}(t_0)\xi^3$. Function $a(t)$ obeys the
Friedman equation
\begin{equation}
\label{dotaa} \left(\frac{\dot a}{a}\right)^2=\frac{8\pi
G}{3c^2}(\varepsilon_r+\varepsilon_m+\varepsilon_{\Lambda}),
\end{equation}
which can be rewritten as $\dot a/a=H_0E(z)$, where redshift $z=a^{-1}-1$,
$H_0$ is the current value of the Hubble constant and function
\begin{equation}
E(z)=[\Omega_{r,0}(1+z)^4+\Omega_{m,0}(1+z)^3+
\Omega_{\Lambda,0}]^{1/2}. \label{efun}
\end{equation}
Here $\Omega_r$, $\Omega_m$ and $\Omega_{\Lambda}$ is the density
parameters (fractions of density to critical density $\rho_c=3H^2/(8\pi G)$)
of radiation (in sum with relativistic neutrino), nonrelativistic matter (DM
and baryons) and $\Lambda$--term, respectively. The $\Lambda$--term affects
the evolution of perturbations only at the redshift $z\leq10$. Relations
$t(z)$ and $z(t)$ can be easily obtained from the solution of the
equation (\ref{dotaa}).

The considered cosmological model is flat:  $\Omega_r+\Omega_m+
\Omega_{\Lambda}=1$. By using the second Friedman equation (for $\ddot
a$) one can rewrite (\ref{d2rdt1}) as follows
\begin{equation}
\frac{d^2b}{dz^2}+\frac{db}{dz}S(z)+
\left(\frac{1+\delta_h}{b^2}-b\right)\frac{\Omega_{m,0}(1+z)}{2E^2(z)}=0,
\label{d2bdz1}
\end{equation}
where function
\begin{equation}
S(z)=\frac{1}{E(z)}\frac{dE(z)}{dz}-\frac{1}{1+z}
\end{equation}
and value of fluctuation $\delta_h=M_h/M_{\rm DM}$. In the limiting
case $\varepsilon_{\Lambda}=0$ the (\ref{d2bdz1}) is equivalent
with the equation, which was obtained at \cite{kt}. Unfortunately,
the fitting formula of \cite{kt} $\rho\simeq140\Phi^3(\Phi+1)\rho_{\rm eq}$,
where $\Phi=\delta_h$, does not allow us to consider the very dense clusters
(with $\delta_h>10^4$) and we must solve Eq.~(\ref{d2bdz1}). In
dependence of epoch one can omit some terms in the r.h.s. of (\ref{d2rdt1}).
We start tracing the evolution of cluster at high redshift $z_i$ when the
considered shell crosses the horizon $r\sim ct$. Initial conditions for
evolution are presented in the Fig.~\ref{massprof1}.

\begin{figure}[t]
\includegraphics[angle=0,width=0.45\textwidth]{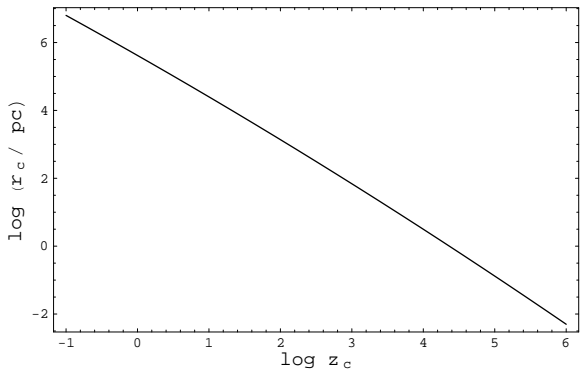}
\caption{\label{solgen1} The virial radius $r_c$ of
galaxy is shown as a function of redshift $z$. }
\end{figure}

The moment of the expansion stop $\dot r=0$ is defined by the condition
$db/dz=b/(1+z)$. We put that after expansion stop. After contraction from
maximum radius $r_s$ to $r_c=r_s/2$ the shell is virialized and fixed at the
radius $r_c$. Therefore, the mean density of the DM in the virialized object 
$\rho$ is 8 times greater then the DM density at the moment of maximum
expansion
\begin{equation}
\rho=8\rho_{m,0}(1+z_s)^3b_s^{-3},
\end{equation}
and an effective (virial) radius of the object
\begin{equation}
r_c=\left(\frac{3M_{\rm DM}}{4\pi\rho}\right)^{1/3}.
\label{rceq}
\end{equation}
The radius (\ref{rceq}) is shown in the Fig.~\ref{solgen1}.

Let us consider the fate of spherical shells by moving our attention from the
center of the cluster to outer regions. It is obvious (and supported by
numerical calculations) that inner more dense shells stop expansion early
than outer shells. As was discussed before, a BH with the mass
$M_c=4.3\cdot10^7M_{\odot}$ forms in the center of the cluster at the moment
$t_i$. The subsequent shells, where $\delta_h>1$ ($M_{\rm DM}<M_h$), are very
dense too. These shells are detached from the cosmological expansion at the
radiation dominated epoch. In general, these shells accreted by the central BH
in the process of two--body relaxation of BHs. This effect will be considered
below. The boundary value $\delta_h=1$ corresponds to the masses
$M_h=M_{\rm DM}=3.3\cdot10^8M_{\odot}$.

\begin{figure}[t]
\includegraphics[angle=0,width=0.45\textwidth]{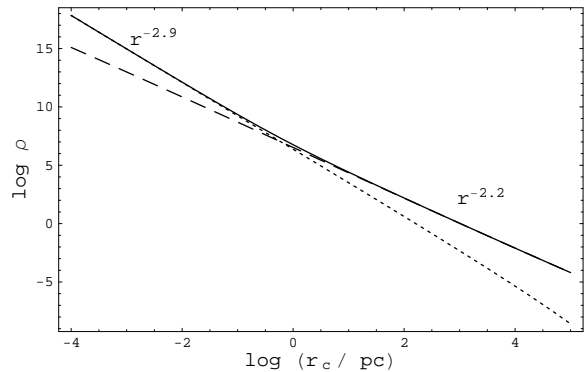}
\caption{\label{massprof2} The final density profile (\ref{dprofeq}) of a
galaxy ($\rho$ in units $M_{\odot}/pc^3$) in dependence of the distance from
the center of cluster $r_c$ is shown for DM (dashed line), for BHs (dotted
line) and for the sum of these densities (solid line). Asymptotic power laws
are also marked. }
\end{figure}

For the outer shells where the fluctuation is small $\delta_h<1$, the known
Meszaros solution is true, according to which the fluctuation growth till the
moment $t_{\rm eq}$ is equal to 2.5. For the early formed PBHs it is possible
the  process analogous to the 'secondary accretion'.  As a result the PBHs
would be 'enveloped' by the DM halo. We call these haloes as induced galaxies
(IG). The density profile does not follow the secondary accretion law
$\rho\propto r^{-9/4}$ because the central mass is not compact. After
virialization the distribution of DM is 
\begin{equation}
\rho_{\rm DM}(r)=\frac{1}{4\pi r^2}\frac{dM_{\rm DM}(r)}{dr},
\label{dprofeq}
\end{equation}
where function $M_{\rm DM}(r_c)$ is determined by the solution of
(\ref{d2bdz1}). In analogy with the DM, one can obtain the profile of
the BHs density and of total density. The corresponding results are shown in
the Fig.~\ref{massprof2}, where density is expressed in units
$M_{\odot}/pc^3$ and distance from the center is in pcs.
As can be viewed from the Fig.~\ref{massprof2}, at the Sun distance from the
Galaxy center $r=8$~kpc the total local mass density is $0.7$~Gev~cm$^{-3}$.
Therefore, some characteristics of the obtained galaxy are similar to the our
Galaxy. But the considered induced galaxy is more dense at its center and
hosting the supermassive BH.

The resultant structure of induced galaxy is the following. Inside radius
$r=10$~pc from the induced galaxy center the mass of BHs (including a central
supermassive BH) is $2.9\cdot10^8M_{\odot}$. Inside the same radius the mass
of DM is $3.4\cdot10^8M_{\odot}$. At the larger distances the
density profile of DM is well approximated by the power low
\begin{equation}
\rho(r)= 2.2\cdot10^4
\left(\frac{r}{10\mbox{~pc}}\right)^{-2.2}\mbox{$M_{\odot}$pc$^{-3}$}.
\label{dmfit}
\end{equation}
The density profile (\ref{dmfit}) differs from the Navarro-Frenk-White
\cite{nfw} and
other proposed profiles, obtained from the numerical simulation of haloes
formation
(without primordial cluster of BHs) but corresponds to these profiles at
intermediate scales where power index $\approx-2$. 
An interesting properties is the diminishing of mean velocity 
$V_v=(GM/2R)^{1/2}$ of IGs in time (with decreasing of $z$).
Such a behavior is a consequence of the shape of perturbation spectrum
produced by
clusters of BHs (see next Section).

\section{Termination of induced galaxies growth }
\label{okonch}

The growth of a virialized region terminates at the epoch of nonlinear growth
of the ambient density fluctuations of the same mass $M$ ( originated from
the standard inflation ) as the our combined system of PBHs plus the DM halo. 
The laws of growth for the both fluctuations at the matter dominated epoch
are the same (\ref{glam}). Therefore, the condition of the growth termination
of a typical induced galaxy is the equality of r.m.s. perturbations
$\delta_{\rm eq}^h(M_{\rm DM})$, produced by the PBHs cluster, and the
inflationary r.m.s. perturbations $\sigma_{\rm eq}^{\rm DM}(M_{\rm DM})$,
both depending on the DM mass scale
$M_{\rm DM}$:
\begin{equation}\label{Deltaeq}
\nu\sigma_{\rm eq}^{\rm DM}(M_{\rm DM})=\delta_{\rm eq}^h(M_{\rm DM}),
\end{equation}
where $\nu$ is the perturbations peak height, in this section we consider only
mean perturbation with $\nu=1$. The statistical properties of the
inflationary Gaussian perturbations are determined by the power spectrum
$P(k)$. We use the power spectrum of the DM from \cite{Dav85}:
\begin{equation}
P(k)=\frac{Ak} {(1+1.71u+9u^{1.5}+u^2)^2},\label{cdm}
\end{equation}
where $u=k/[(\Omega_{m,0}+\Omega_{b,0})h^2\mbox{~Mpc}^{-1}]$, and
$k$ is the comoving wave vector in  Mpc$^{-1}$ units. The initial spectrum is
supposed to be the Harrison--Zeldovich spectrum. The normalizing constant $A$
corresponds to the value $0.9$ of r.m.s. fluctuations at  8~Mpc scale. The
r.m.s. perturbation at the mass scale $M$ corresponding to the radius $R$ is
\begin{equation}
\sigma^{\rm DM}(M(R))=\frac{1}{2\pi^2}\int\limits_0^\infty
k^2\,dk\,P(k)W(k,R), \label{sig}
\end{equation}
where $W(k,R)$ is a filtering function. The evolution of the cosmological
perturbations in the universe with the $\Lambda$--term at the matter dominated
epoch can be obtained from (\ref{d2bdz1}) or simply from the equation
\cite{mo02}
$\delta(t)/\delta(z_{\rm eq})=g(z)(1+z_{\rm eq})/[g(t_{\rm eq})(1+z)]$, where
\begin{equation}
g(z)\approx\frac{5}{2}\Omega_m[\Omega_m^{4/7}-\Omega_{\Lambda}+
(1+\Omega_m/2)(1+\Omega_{\Lambda}/70)]^{-1}, \label{glam}
\end{equation}
where $\Omega_m=\Omega_{m,0}(1+z)^3/E^2(z)$ and so on.

\begin{figure}[t]
\includegraphics[angle=0,width=0.45\textwidth]{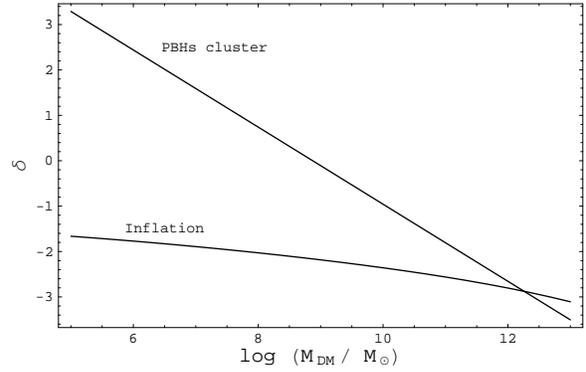}
\caption{The r.m.s. density perturbation values at the time $t_{\rm eq}$ of
matter--radiation equality produced by the cluster of PBHs and by the
inflation respectively.}
\label{delfig} 
\end{figure}

The value of fluctuations produced by the cluster $\delta_{\rm eq}^h(M_{\rm
DM})=2.5\delta_i(M_{\rm DM})$ is shown in the Fig.~\ref{delfig} together with
the Harrison-Zeldovich spectrum (\ref{cdm}) inserted to (\ref{sig}).
Fluctuations produced by the cluster are very well approximated by the power
law \begin{equation}
\delta_{\rm eq}^h(M)=0.044
\left(\frac{M}{10^{10}M_{\odot}}\right)^{-0.85}. 
\label{dhm}
\end{equation}

From the numerical solution of (\ref{Deltaeq}) we obtain the mass of the DM
halo at the moment of growth termination $M_{\rm
DM}=1.8\cdot10^{12}M_{\odot}$. The expansion stops at the redshift $z=1.64$.
Note that growth termination does not mean the termination of large scale
structure formation. This means only the termination of the contraction under
the influence of PBHs cluster. After this moment the structure formation
proceeds by the standard hierarchical clustering scenario: galaxies are
assembled into larger ones, clusters and superclusters. Therefore, the large
scale structures in the our model is the same as in the standard cosmological
scenario.

The formed induced galaxies looks like the giant elliptical galaxies with a
small ellipticity. They have the central DM spike shown in the
Fig.~\ref{massprof2} and central supermassive BH.


\section{Evolution of quasar activity}

In view of the discovery of distant quasars with the redshifts $z>6$, let us
discuss the properties of induced galaxies at these redshifts.  The described
model of induced galaxies can easily explain the $M_Q=3\cdot10^9M_{\odot}$
supermassive BHs at $z=6$ (the age of the universe $t_6=9\cdot10^8$year).  In
this model there are supermassive BHs formed at the radiation dominated epoch
long before the galaxies formation.  Initially there is at least
$M_c=4.3\cdot10^7M_{\odot}$ or even larger (if there is an adsorbtion of
surrounding smaller BHs from cluster by the most massive one) BH at
the cluster center.  Let us suppose that this BH radiated at the Eddington
limit, $L_E=1.3\cdot10^{46}$erg/s.  This BH grows exponentially with a
characteristic time $t_E=4.5\cdot10^8\eta$~yrs, where $\eta$ is the
effectiveness of the accretion matter--energy transformation.  The time of
growth from $M_c=4.3\cdot10^7M_{\odot}$ to $M_Q=3\cdot10^9M_{\odot}$ is
$\Delta t=t_E\ln{(M_Q/M_c)}\simeq2\cdot10^8$year for $\eta\sim0.1$.  This
means that QSO activity began at the moment $t_6-\Delta
t\simeq7\cdot10^8$~yr, corresponding to the redshift $z=7.3$.  At this
redshift the induced galaxies have the virial mass
$4.3\cdot10^{11}M_{\odot}$, the virial radius $40$kpc, the mean virial
velocity $140$~km/s and the mean virial temperature 
$T_v=m_pv^2/3\simeq2\cdot10^6$~K, where $m_p$ is the proton mass.  These
values were obtained from the numerical solution of the equation
(\ref{d2bdz1}).  For these parameters the radiative cooling
mechanism for the gas in the induced galaxies is effective.  This provide the
cool gas flows necessary for the accretion.

The very popular model of quasars ignitions is the gas flow onto the central
black hole due to the tidal interactions during the merging of galaxies.  In
our model the huge tidal interactions are produced at the time of termination
of the induced galaxies growth considered in the preceeding Section.  In
general, the density distribution of surrounding inflationary perturbations
are not spherically symmetric themselves and with respect to the induced
galaxy center.  This causes the strong tidal forces, instabilities and gas
supply to the central region necessary for the effective accretion and quasar
activity.  We expect the maximum activity at the time of the termination of
the induced galaxies growth at $z\sim1.6$.  This epoch is in agreement  with
the observable shape of quasar redshift distribution.

\begin{figure}[t]
\includegraphics[angle=0,width=0.45\textwidth]{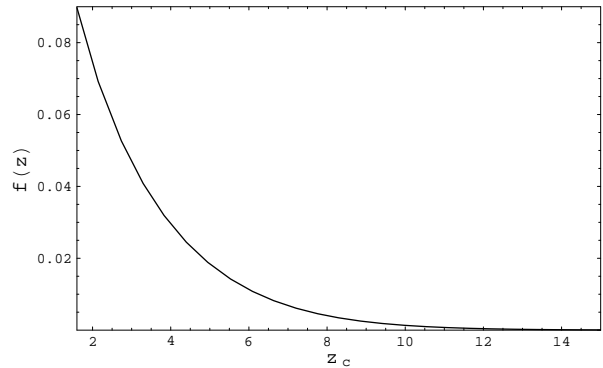}
\caption{The form of quasar ignition events distribution beyond the redshift
$z=1.6$}
\label{qsofig} 
\end{figure}

For the more detailed description of induced galaxies formation it is
important to take into account distributions both the induced galaxies (or
initial PBHs clusters) and inflation perturbations. For simplicity, let us
consider only the Gaussian distribution of inflation perturbations over 
$\nu$.  In a mean $\nu=1$, the fluctuations produced by BHs cluster till
$z=1.6$ prevail over the standard inflation perturbations.  But there are the 
less common fluctuations with $\nu>1$ which exist at the discussed redshifts
$z=6$, $7.3$ and larger. These rare fluctuations responsible in the our model
for the tidal forces and quasar ignition at large $z$.  From the numerical
solution of (\ref{d2bdz1}) we can obtain the virial mass of induced galaxies
in dependence of the redshift $M_{\rm DM}(z)$. Substituting this function
into (\ref{Deltaeq}) we get the dependency over $\nu(z)$.  The $\nu(z)$ is the
required peak height of the inflationary fluctuation for the termination of
the induced galaxies growth and quasar ignition. The distribution of
probabilities of these events is calculated as 
$$
f(z)=-\frac{d}{d
z}\int\limits_{\nu(z)}^{\infty}\frac{1}{\sqrt{2\pi}}e^{-\nu^{'2}/2}d\nu'=
\frac{1}{\sqrt{2\pi}}e^{-\nu^{2}(z)/2}\frac{d\nu(z)}{dz}
$$
and shown in the Fig.~\ref{qsofig}.
This Figure shows the distribution law of the quasar ignition events beyond
$z=1.6$ where the hierarchical clustering begins. Note, that the
Fig.~\ref{qsofig} doesn't present quasar distribution
function $F_{\rm QSO}(z)$. To calculate $F_{\rm QSO}(z)$ we need the poorly
defined  quantity --- the time of quasar activity $t_q(z,M_{\rm BH})$ (see
\cite{DEO} for details). For a given $t_q(z,M_{\rm BH})$ it is possible to
calculate $F_{\rm QSO}(z)\simeq \delta z(t_q,z)f(z)$, where $\delta z(t_q,z)$
is the redshift interval corresponding $t_q$ at redshift $z$.

\section{Conclusion}

In this paper we describe the new model of quasar formation initiated by the
earlier formation of the cluster of PBHs.  This model provides the early
formation of large galaxies with the massive BHs in their
centers.  Nowadays these galaxies could be seen as distant quasars.  The main
calculated parameters of typical galaxy are the following:  the galaxy mass 
$2\cdot10^{12}M_{\odot}$, the central BH mass $4\cdot10^7-10^9M_{\odot}$. The
density profile of these induced galaxies is a near isothermal, $\rho\propto
r^{-2.2}$.  The induced galaxies could be the cause of the of early quasar
activity.  The model predicts the rapid decay of the quasar activity at
redshifts $z\geq10$.  The alternative scenario could be based on the
formation of the less massive PBHs clusters.  In this case each nowadays
galaxy contains numerous BHs originated from the dwarf induced galaxies
aggregated in a single galaxy in the process of hierarchical clustering.  This
possibility will be considered in a separate paper.

\Acknow{
V.I.D and Yu.N.E. were supported in part by the Russian Foundation for
Basic Research grants 02-02-16762-a, 03-02-16436-a and
04-02-16757-a and the Russian Ministry of Science grants
1782.2003.2 and 2063.2003.2. S.G.R.  was supported
in part by the State Contract 40.022.1.1.1106,  by the RFBR grant
02-02-17490 and grant UR.02.01.008.  (The University of Russia). 
            }

\end{document}